
\magnification=\magstep1
%
\font\tenss=cmss10
\font\sevenss=cmss10 at 7pt
\font\fivess=cmss10 at 5pt
\font\tenssi=cmssi10

\textfont0=\tenss \scriptfont0=\sevenss \scriptscriptfont0=\fivess
\newfam\itfam  
\textfont\itfam=\tenssi
\newdimen\aadimen
\def\AA{\leavevmode\setbox0\hbox{h}\aadimen\ht0\advance\aadimen-1ex\setbox0
        \hbox{A}\rlap{\raise.67\aadimen\hbox to \wd0{\hss\char'27\hss}}A}
%
\font\text=cmss10
\font\bold=cmssbx10
\font\italic=cmssi10
\font\bigbold=cmssbx10 at 14.4pt
\font\bigtext=cmss10 at 12pt
\font\minibold=cmssbx10 at 8pt
\font\minitext=cmss10 at 8pt
%
\def\MAINTITLE#1{\footline={\hfill}\nopagenumbers\bigbold\parindent=0pt\hrule\vfill\par\advance\baselineskip by 7pt\par #1\par\advance\baselineskip by -7pt\par}

\def\MANYAUTHORS#1#2{\vfill\bigtext\hrule\vfill#1\vskip2mm#2}
\def\INSTITUTE#1{\vfill\parindent=1em\text\hrule\vfill#1}
\def\MAKETITLE#1{\vfill\parindent=0pt\hrule\vfill#1\vfill\hrule
    \eject\footline={\hfill\folio\hfill}\pageno=1}
\def\ABSTRACT#1{\pageno=1\parindent=20pt\bold\noindent Abstract.
    \text\vskip12pt\noindent#1\vfill\eject}
\def\TITLEA#1{\vskip24pt\bold\noindent #1\vskip12pt\text\noindent}
\def\TITLEB#1{\vskip12pt\italic\noindent #1\vskip6pt\text\noindent}

\def\FIGURE#1#2{\par\advance\baselineskip by -2pt\noindent\minibold Figure
    #1.\ \minitext #2\par\advance\baselineskip by 2pt\par\text}
\def\SIDEFIGURE#1#2#3#4{\midinsert\par\advance\baselineskip by -2pt
    \par\hangindent=#2\hangafter=0\noindent\minibold Figure #3.\ \minitext
    #4\par\advance\baselineskip by 2pt\par\text\vskip#1}

\def\TABCAP#1#2{\par\advance\baselineskip by -2pt\par\noindent\minibold Table
    #1.\ \minitext #2 \par\advance\baselineskip by 2pt\par\text\vskip2mm}

\def\BEGREF{\vfill\eject\TITLEA{References}}
\def\REF{\par\hangindent=20pt\parindent=0pt\hangafter=1}
\def\ENDREF{\parindent=20pt}
\def\la{\mathrel{\mathchoice {\vcenter{\offinterlineskip\halign{\hfil
    $\displaystyle##$\hfil\cr<\cr\noalign{\vskip1.5pt}\sim\cr}}}
    {\vcenter{\offinterlineskip\halign{\hfil$\textstyle##$\hfil\cr<\cr
    \noalign{\vskip1.0pt}\sim\cr}}}
    {\vcenter{\offinterlineskip\halign{\hfil$\scriptstyle##$\hfil\cr<\cr
    \noalign{\vskip0.5pt}\sim\cr}}}
    {\vcenter{\offinterlineskip\halign{\hfil$\scriptscriptstyle##$\hfil
    \cr<\cr\noalign{\vskip0.5pt}\sim\cr}}}}}
\def\ga{\mathrel{\mathchoice {\vcenter{\offinterlineskip\halign{\hfil
    $\displaystyle##$\hfil\cr>\cr\noalign{\vskip1.5pt}\sim\cr}}}
    {\vcenter{\offinterlineskip\halign{\hfil$\textstyle##$\hfil\cr>\cr
    \noalign{\vskip1.0pt}\sim\cr}}}
    {\vcenter{\offinterlineskip\halign{\hfil$\scriptstyle##$\hfil\cr>\cr
    \noalign{\vskip0.5pt}\sim\cr}}}
    {\vcenter{\offinterlineskip\halign{\hfil$\scriptscriptstyle##$\hfil
    \cr>\cr\noalign{\vskip0.5pt}\sim\cr}}}}}
\def\dot{\mathaccent"70C7}

\MAINTITLE{Our Galactic Center:  A laboratory\hfill\break for the feeding
           of AGNs?}

\MANYAUTHORS{Susanne von Linden$^{\rm 1}$, Peter L.\ Biermann$^{\rm 1}$,
             Wolfgang J.\ Duschl$^{\rm 1,2,3}$,}{Harald Lesch$^{\rm 1}$,
             Thomas Schmutzler$^{\rm 2,4}$}

\INSTITUTE{
\item{1:}Max Planck Institut f\"ur Radioastronomie, Auf dem H\"ugel
69, Postfach 2024,\hfill\break D-53010 Bonn, Germany
\item{2:}Institut f\"ur Theoretische Astrophysik, Universit\"at Heidelberg, Im
Neuenheimer Feld 561, D-69120 Heidelberg, Germany
\item{3:}Interdisziplin\"ares Zentrum f\"ur Wissenschaftliches Rechnen,
Universit\"at Heidelberg,\hfill\break Im Neuenheimer Feld 368, D-69120
Heidelberg, Germany
\item{4:}Max Planck Institut f\"ur Kernphysik, Saupfercheckweg 1, D-69117
Heidelberg,
Germany}
\MAKETITLE{accepted for publication in \italic Astronomy and Astrophysics}
\ABSTRACT{
We demonstrate that our Galactic Center, despite little evidence for the
presence of a currently active nucleus, provides insight into the feeding of
AGN:  The
observed velocity field of molecular clouds can be interpreted as tracing out
the spiralling inwards of gas in a large accretion flow towards the Galactic
Center (Linden et al. 1993, Biermann et al. 1993) in the radial distance range
from a few parsec to a few hundred pc.
The required effective viscosity corresponds well to the
observed turbulent velocities and characteristic length scales. The implied
mass influx provides indeed all the material needed to maintain the presently
observed star formation rate at distances closer than about $100$ pc.  We
argue that the energy input from supernova explosions due to the high rate
of star formation can feed the turbulence of the interstellar medium.
This then keeps the effective viscosity high as required to feed the star
formation.  We suggest that this process leads to limit cycles in star
formation,
and as a consequence also to limit cycles in the feeding of any activity at the
very center.}

\TITLEA{1.\ Introduction}%
The feeding of active galactic nuclei, often of a power that exceeds all of the
stars of the surrounding galaxy put together, is one of the major riddles in
understanding the energetics of quasars, Seyfert galaxies, BL Lac nuclei and
other forms of AGN.  It has been argued convincingly that this requires a major
mechanism to remove the angular momentum of gas which is already inside several
hundred parsec of the nucleus.  With the speed of sound at the observed gas
temperature and the size of the clouds as scaling for the turbulent viscosity
to
drive an accretion disk, this does not work; therefore, Shlosman et al.\ (1989)
argued that a stellar bar is required to give the torque which removes angular
momentum.  While bars are indeed observed in many galaxies, and visibly
interact
with the gas (e.g.\ NGC$\,$1300), this is by no means clear for all galaxies
that harbor an active nucleus.  Also, the bars observed do not easily solve the
angular momentum problem on the small length scales required.  The problem is
actually worse, since many galaxies have an inner region which has a large
amount of molecular gas with a large rate of star formation, which also needs
to be fed.  Shlosman et al. (1990) argue that there is a possibility of ``hot
accretion", which would indeed provide larger accretion rates, but their model
does not easily yield to a detailed observational check.

Here we will explore our Galactic Center to see what we can learn about the
mass
influx in an environment where there is little signature of an AGN, and which
therefore provides a baseline for which physical concepts can be tested also
for
inner regions of galaxies that do harbor an active nucleus quite clearly.

Recently we demonstrated (Linden et al. 1993, Biermann et al. 1993) that the
observed velocity field of several clouds at distances from the Galactic
Center of near $10$ pc to\ $\approx 100$ pc can be interpreted as tracing out a
large
scale accretion flow.  The fit to the observational data implies an effective
viscosity for the transport of angular momentum outwards and gas inwards.  This
effective viscosity in fact matches approximately that value for the transport
coefficient implied by the overall velocity dispersion of the clouds and their
scale height; this modelling also implies a mass flow inwards, and indeed, the
mass flow in turn approximately matches the mass flux required to sustain the
presently observed star formation rate.

Moreover, the mass flow rate is in good agreement with what one deduces from
observations of the Circumnuclear Disk (CND) inside the region Linden et al.
and Biermann et al. were using (Genzel and Townes 1987, Jackson et al. 1993).

We propose that the energy input from supernova explosions due to a
high rate of star formation can feed the required turbulence of the
interstellar medium.
Therefore we specifically ask in this paper whether the star formation rate
can uphold the energy dissipation required to maintain the convective
turbulence observed, and thus ask, whether the causal loop can be closed
between
high star formation rate, high excitation of turbulent convection, thus high
effective viscosity, thus high mass influx, and then in turn, high star
formation
rate.
\vfill
\eject
\TITLEA{2.\ The velocity distribution of the clouds}%
The observations of Zylka (1990) and Zylka et al.\ (1990) of molecular line
emission in the central region of our Galaxy provide the dataset from which we
will start to raise our argument.  These data allow for the first time to
unambiguously discuss the dispersion of the cloud motion and also allow first
attempts to actually locate some of the larger clouds, giving the impression of
a moderately thick disk-like distribution.

The velocity spread of the cloud motions is
about $50\,\rm km/s$ and the scale height of their distribution is of the order
of $10\,\rm pc$.  In fact, we can make the following check: Is the cloud
velocity dispersion that implied by the known gravitational potential and the
$z$-distribution of clouds,
which would support the notion that the cloud motion is
basically isotropic?  These clouds cannot be moving in an external medium at
supersonic speed (referred to the outer medium), because they would be
destroyed
on dynamical timescales. Therefore, the cloud motions must reflect the motions
of a hotter gas, of which the sound speed is at least the velocity of the
clouds, and of which the scale height again is at least that of the cloud
distribution. In fact, the solar neighborhood suggests that the scale height
of the hot ionized medium is indeed several times larger than that of the
molecular cloud layer (Garcia-Munoz et al. 1977, Reynolds 1989, 1990).

In order to estimate the scale height of the gas distribution,
we have to specify the gravitational potential in which the disk is
situated. The mass
inside a spherical radius $r$ can be approximated (Sanders \& Lowinger 1972,
Genzel \& Townes 1987) as

$$M_r=M_0\,(r/r_0)^{5/4} \eqno{1}$$
with $M_0=5.8\,10^7\,\rm M_{\odot}$
and $r_0=9.2\,\rm pc$, which is exactly $30$ lightyears.  This implies from
hydrostatic equilibrium (assuming the temperature to be constant in $z$) a
scale
for the hot outer gas of

$$H=0.61\,{\rm pc\ }\left({{c_s} \over {10\,{\rm km/s}}}\right)\,(s/r_0)^{7/8}
\;,\eqno{2}$$
where $s$ is the radial distance from the Galactic Center
in the disk, and $c_s$ denotes the sound speed. Using, say, $s=50\,\rm pc$ and
$H=10\,\rm pc$ from the Zylka et al.\ data, we obtain an adiabatic speed of
sound of $37.3\,\rm km/s$ which corresponds to a temperature of $6.1\,10^4\,\rm
K$.
This is surprisingly close to the temperature which characterizes the radiation
field of Sgr A* (see the recent discussion by Falcke et al.\ 1993). We have
thus
made a consistency check: The cloud velocity dispersion in the $z$-direction is
to within the errors the same as along the line of sight, and so the cloud
velocity distribution is approximately isotropic.

\TITLEA{3.\ The physical cause for the large velocity spread of clouds}%
\TITLEB{3.1\ Global gravitational instability}%
Lin and Pringle (1987a) have argued that in a self-gravitating disk the overall
gravitational instability serves to excite motions which lead then to an
effective turbulence and to a turbulent momentum transport. This mechanism
may serve as initial viscosity injection into an interstellar medium where
star formation just begins. The gravitational instability in a disk (Toomre
1964) appears if

$$Q={v_d \kappa\over{\pi G\Sigma}}<1 \eqno{3}$$
Here $v_d$ denotes the velocity dispersion, $\kappa$ is the epicyclic
frequency, and $\Sigma$ is the gas surface mass density.
Equation 3 translates to

$$v_d = 1.4 \,{\rm km/s\ } (s/r_0)^{-1/8} \, Q \,, \eqno{4}$$
using already here the radial surface density dependence discussed below.

The gravitational instability leads to the growth of density perturbations and
can cause a rotating disk to fragment into clumps. The angular momentum
transport
is then provided by the shear and viscosity produced by this kind of
instability.
The viscosity is related to the clouds that form in a gravitationally unstable
disk. The instability develops fragments with velocities of the order

$$v_c={\pi G\Sigma\over{\kappa}} \eqno{5}$$
and length scales

$$\lambda={2\pi^2 G\Sigma\over{\kappa^2}} \eqno{6}$$
We assume that initially the galactic disk is kept on the border of
gravitational instability, creating a cloudy medium with both cloud
sizes and separations of order $\lambda$. These clouds should provide
the angular momentum transport by a 2-dimensional random walk with step-length
$\lambda$ and velocity $v_c$, i.e. the viscosity is given by (Lin and Pringle
1987b)

$$\nu_{\rm eff}\simeq \lambda v_c \simeq \lambda^2 \Omega(s), \eqno{7}$$
where $\Omega(s)$ is the rotation frequency of the disk.
\par
The onset condition of the instability $Q<1$ defines a critical surface mass
density $\Sigma_c$. Kennicutt (1989) presented overwhelming evidence, that in
15 disk galaxies the star formation threshold is associated with the onset of
gravitational instabilities in the gas disk. His sample shows that for
radii smaller than 5 kpc $\Sigma >\Sigma_c$, so that $Q<1$ and the
viscosity may be given by $\nu_{\rm eff}$.

While this, in principle, could be a viable way of transporting
angular momentum and mass, in practice, it turns out not to be
efficient enough compared to the  observed flow velocities in the innermost
few hundred parsec of galactic centers.
\TITLEB{3.2\ Supernovae driven turbulence -- the basics}%
Related to the described scenario is another obvious possibility:  General star
formation, driven by gravitational instabilities, leads also to
the formation of massive stars, which in turn not only have strong winds
(e.g. Bieging et al.\ 1989), but also produce supernova explosions.  The energy
input from these supernovae is for massive stars similar to the wind energy
input integrated over the lifetime of the star. Both together can put
sufficient
energy into an interstellar medium not just to stir it, but even to blow it up:
This effect is clearly seen in the starburst galaxy M82 (Kronberg et al.\ 1985,
Schaaf et al.\ 1989), where the interstellar medium is blown out to distances
of
many kpc perpendicular to the central part of the galactic disk (see various
reports in Bloemen 1991).  A fair
number of galaxies are now known to behave similarly.  Hence it is
evident that
massive stars provide a strong stirring.  The question here is why the stirring
by
massive stars would produce larger velocities near to the Galactic Center than
in
the solar neighborhood.
We propose to ask the question in the same framework as the model of McKee and
Ostriker (1977) for the interstellar medium in the solar neighborhood.
Specifically we aim at a consistency check for the notion proposed, i.e., that
the high star formation rate can uphold the high rate of turbulence.

First of all, we wish to demonstrate that the star formation rate is very much
higher in the Galactic central region:  The star formation rate is observed to
be near $0.5 \, M_{\odot}$/yr in the radial range to a few hundred pc (G\"usten
1989) , which is $300$ times the rate of star formation per area over the
wholeGalaxy, with a hard lower limit at about $0.05\,M_\odot$/yr.
Since the scale height near 100 pc from the Galactic Center is
about $10$ times smaller than in the solar neighborhood,
the implied supernova rate per unit of time and unit
volume is about $3 \, 10^3$ times higher.  Thus, scaling the
supernova rate with the star formation rate, the supernova rate is about
$S_{-13}
\, = 3 \, 10^3$, where $S_{-13}$ is the supernova rate per
volume in units of $10^{-13}  \, \rm pc^{-3} \, yr^{-1}$, an estimate which we
will reduce by half an order of magnitude
to be conservative.  We will use this
final estimate below. However, the density of the interstellar medium and thus
the
pressure is also higher (for a recent investigation of star formation near the
Galactic Center, see Morris 1993).

Hence, second, we wish to demonstrate that the filling factor of the
large number of supernova remnants inside the interstellar medium is of order
unity, which would imply that the turbulence generated by the supernovae
indeed reaches most of the
medium.  Only for a filling factor
of order unity are all the supernovae capable of
stirring up the interstellar medium fairly completely.  The expansion of a
supernova shell can be divided into an adiabatic phase until cooling sets in,
and then a second phase of constant momentum.  The various pieces of the
former
shell keep expanding to a radius $R_E$, when their kinetic energy approximately
matches the average pressure of the medium (Chevalier 1974, McKee and Ostriker
1977).  We write the ambient pressure as $P_{0,4k} \, = \, 10^{-4} \, P_0 \,
{\rm dyn}/k$, the supernova energy input as $E_{51} \, = \, 10^{-51} E_{\rm SN}
\rm \, erg$, and the ambient intercloud density as $n_0\, \rm particles/cm^3$;
below we will always use $E_{51} \, = \, 1$.  We then have

$$R_E \, =\, 55 \, {\rm pc} \, E_{51}^{0.32} \, n_0^{-0.16} \, P_{0,4k}^{-0.20}
,\eqno{8}$$
from a parametrization of Chevalier's
(1974) results given by McKee and Ostriker
(1977). The cavity survives until the surrounding gas encroaches upon it, for a
time $t_{E}$, given by

$$t_E \, =\, 7 \, 10^6 {\rm yrs} \, E_{51}^{0.32} \, n_0^{0.34} \,
P_{0,4k}^{-0.70}.\eqno{9}$$

Here we have to estimate the likely range of the intercloud  density and
temperature.  From the scale height and the known gravitational potential we
can
estimate the intercloud temperature to be near $10^6 \, \rm K$.  X-ray data
(Blitz et al.\ 1993, Sunyaev et al.\ 1993, Markevitch et al.\ 1993)
suggest that the temperature might even be higher, with a range up to about
$10^8$ K, at an associated density of the tenuous gas of near $0.05$
particles/cm$^3$; however, at such a temperature there has to be
a strong wind out of the Galaxy which is not supported by other evidence, and
hence we will use the lower estimate for the temperature, a range of $10^{6 \pm
1} $ K.  With an average density of interstellar gas of $240$ particles/cm$^3$
out to about $500$ pc (G\"usten 1989) , using an estimated thickness of the
cloud
layer there of near $30$ pc (Zylka et al. 1990), and a filling factor of the
clouds of $f_{cl} \la \,0.1$,
we can estimate a cloud density. We assume
approximate pressure equilibrium between clouds, of an observed temperature in
the
range $40 \, {\rm to} \, 100$ K, and intercloud medium. We thus estimate the
intercloud density to about $0.01 \, {\rm to} \, 1 \, f_{cl,-1}^{-1}\rm
particles/cm^3$.
{}From the observed strengths of the magnetic fields (Blitz et al.
1993, Morris 1993) we obtain an independent estimate of the average
pressure of order $3 \, 10^6$ in units of  ${\rm dyn}/k$, which corresponds to
an intercloud density of about $1$ particles/cm$^3$ at $T=10^6$ K.  In summary
there is evidence for
a hot intercloud gas with an estimated pressure of $P_{0,4k} \, = \, 300$, and
a
density in the range of $0.01$ to $10$ particles/cm$^3$.
This leads to a maximum
expansion radius in the range of $12$ to $40$ pc, and a maximum time to caving
back in for the bubble in the range of about $3 \, 10^4$ to $3 \, 10^5$ yrs.
The
chance for a given point to be in a supernova remnant is given by

$$Q_{\rm SNR} \, =\, 0.5 \, \, S_{-13} \,E_{51}^{1.28} \, n_0^{-0.14} \,
P_{0,4k}^{-1.30},\eqno{10}$$
while the filling factor is given by

$$f_{\rm SNR} \, = \, 1 - e^{-Q_{\rm SNR}} .\eqno{11}$$
Putting in our estimates for the possible range of numbers we obtain for
$Q_{SNR}$
an estimated range of $0.2$ to $0.6$, and for the filling factor of supernova
remnants $0.2$ to $0.5$.
The uncertainty in our understanding of the
interstellar medium near the Sun is substantial and even more so near the
Galactic Center, and so these numbers have to be used with some caution. Also,
all these numbers should depend on the radial distance $r$; here we just
estimated the relevant orders of magnitude and showed
that it is actually worthwile to go through the calculation at all.

We thus conclude that it is possible that Sedov type supernova explosions can
indeed reach a sufficient fraction of the interstellar medium even near to
the Galactic Center to domiante the turbulent motion.

\TITLEB{3.3.\ Supernovae driven turbulence -- the details}%
Consider the explosion of a star into a homogeneous interstellar medium (Cox
1972) of density $n_0$. The explosion of a star into the interstellar medium
can be modified strongly by cloud evaporation and many other effects, discussed
in detail by McKee and Ostriker (1977).  However,
kinematic investigations of the supernova remnants Tycho (Strom et al. 1982),
Kepler (Dickel et al. 1988), and SN1006 (Long et al. 1988) show that while
there are deviations from the ideal d log $r$ / d log $t$ = $2/5$ law,
they are not substantial.
Therefore, we shall adopt the simple Sedov expansion as our model.  The
adiabatic Sedov expansion begins to be modified when cooling sets in.  Allowing
for
time dependent cooling and new atomic rates, Schmutzler and Tscharnuter (1993)
demonstrate that the approximation used by Cox for the cooling coefficient
$\Lambda=10^{-22}\,{\rm erg\, cm^3/s\ }n_0^2 = \,L\,n_0^2$
is sufficient for our purposes; in fact,
their
calculations show that this cooling coefficient is good to within a factor of 2
in the temperature range from $10^6$ to $5\,10^6\, \rm K$ and even a fair
approximation at lower temperatures above $2 \, 10^4 \rm K$. So we have for the
radius of the supernova shell $r_{\rm SN,cool}$ at the point when cooling
begins
to affect the shell
$$r_{\rm SN,cool}=20.9\>{\rm pc\ }E_{51}^{3/11}\,(10^{22}L)^{-2/11}\,
n_0^{-5/11}\;, \eqno{12}$$
where $E_{51}$ is the explosion energy in units of $10^{51}\,\rm erg$. The
velocity of expansion at that point is given by

$$v_{\rm SN,cool}=281\>{\rm km/s\ }E_{51}^{1/11}\,(10^{22}L)^{3/11}\,
n_0^{2/11}\;. \eqno{13}$$

Already the statistical work of Berkhuijsen (1986) has demonstrated that the
dependence on the environmental density $n_0$ is one of the overriding effects
in supernova expansion and leads to strong selection effects.

Here we note that the expansion begins to be affected by cooling and still
shows a fairly high velocity of expansion at a radius which is only weakly
influenced by the environmental density, so that the higher density in the
hotter medium outside
the clouds near the Galactic Center as compared to the solar neighborhood has
only a moderate influence.  It appears reasonable to suppose that the velocity
at
which the interstellar medium turbulence gets fed, is the velocity at which the
supernova shells abruptly start cooling, because then these shells become
unstable,
disrupt and fly apart in various cloud fragments.  We note again, that the
expansion will actually continue much further, with the various parts of the
shell coming to rest when the expansion slows down to the average
characteristic
velocity of the medium, the effective speed of sound.

In McKee\&Ostriker's picture, the expansion is limited by reaching the
local speed of sound which is assumed to be given. Here, in contrast, we
assume that the expansion sets the speed of sound in the tenuous environment
and that the limiting condition is rather given by the dissipation
requirement (see sect.\ 5, and, especially, eq.\ 29).

Let us suppose, that this is the case and identify the velocity of the
supernova
shell with the velocity  dispersion of clouds to within some factor $f_{\rm
dis}$, left
unspecified for the moment, but expected to be much less than unity (otherwise,
there would not be any shock). We also suppose that the intercloud medium is in
pressure equilibrium with the internal cloud pressure.  G\"usten (1989) gives
the integrated surface density out to $500\,\rm pc$ as in the range $100\,\rm
M_{\odot}\,pc^{-2}$ to $400\,\rm M_{\odot}\,pc^{-2}$.  Since there are
ununderstood discrepancies with $\gamma$-ray emission (Lebrun et al. 1983,
Bloemen et al. 1984, G\"usten 1989) we prefer to err on the cautious side and
will
use the lower limit in the following. Observations show that the gas surface
density decreases with distance from the Galactic Center, and we adopt here
somewhat arbitrarily an $s^{-1}$ law.  We will show further below that a
moderately steep powerlaw like this is a natural dependence.  All this then
means
that we adopt in the following

$$\Sigma =2.5\,10^{23}\>{\rm atoms/cm^2\ }\,(s/r_0)^{-1}\; .\eqno{14}$$
Then we have the following relations: First the relation between the column
density, the scaleheight and the average density -- dominated by the clouds.
This average density, being equal to the cloud density to within a volume
filling factor $f_{\rm cl}$, say, of order 0.1, then by pressure equilibrium
gives an
intercloud density, which in turn enters the expansion law for the shell, which
again gives a characteristic velocity which we then -- and this closes the loop
-- set proportional to (but $\gg$ than) the intercloud speed of sound and the
velocity dispersion of the clouds. Hence the first relation is

$$n_{\rm cloud}=2.5\,10^{23}\,{\rm atoms/cm^3\ }f_{\rm cl}^{-1}\,H^{-1}\,
(s/r_0)^{-1}\;.\eqno{15}$$
Pressure equilibrium gives the second relation

$$n_{\rm intercloud}=(25/21)\,c_s^{-2}\,n_{\rm cloud}\,c_{s,\rm cloud}^2
\eqno{16}$$
where we use the appropriate adiabatic constant for the two atomic molecular
and the single atomic gas.
We identify here $c_s$ with the intercloud speed of sound, and the
intercloud density with $n_0$, on the grounds that supernova expansion into the
intercloud medium alone can excite overall turbulence, while an explosion into
a dense cloud is quickly stifled. We assume -- and this has little consequence
--
that the speed of sound inside the clouds is of order $0.76\,\rm km/s$,
corresponding to an adopted temperature inside the cloud of $100\,\rm K$.
The final relation to close the loop is

$$c_s=v_{\rm SN,cool}\,f_{\rm dis},\eqno{17}$$
where we have to use the intercloud density in the expression for the
expansion.
Surely, this factor $f_{\rm dis}$ is small, since the fragment velocities get
further
dissipated, say, to $f_{\rm dis}$ of order 0.1. Combining all these equations
yields a
relation for the radial behaviour of the implied speed of sound and its value,
which we can then again compare with observations

$$c_s=56\>{\rm km/s\ }C\;(s/r_0)^{-15/68}\eqno{18}$$
with

$$\eqalign{C\;=\;&E_{51}^{1/17}\,(10^{22}L)^{3/17}\,(f_{\rm dis}/0.1)^{11/17}\,
(f_{\rm cl}/0.1)^{-2/17}\times\cr
&\times(c_{s,\rm cloud}/ 0.76\,\rm km\,s^{-1})^{4/17}}\;. \eqno{19}$$
We note that the result depends only with the power 2/17 on the cloud
temperature.  We emphasize that the parameters $f_{\rm dis}$ and $f_{\rm cl}$
have been chosen
to be of order $0.1$.  In reality, their value could be different, e.g.,
smaller,
and obviously, their value likely depends on radial distance $r$ in the Galaxy.
However, the dependence of the parameter $C$ is only rather weak on these two
parameters; for $f_{\rm dis}=0.03$ we still get a value for $C$ only a factor
of $2$
lower, and for $f_{\rm cl}$ = $0.01$ we get a value of $C$ = $1.3$ times
larger.  To
test it
with $f_{\rm dis}=0.1$ and $f_{\rm cl}=0.1$ we calculate first the value that
this relation
gives
for a radius in the disk of $50\,\rm pc$, which is $39.4\,C\,\rm km/s$, which
is
to be compared with the implied speed of sound from the scale height of
$37.3\,\rm
km/s$ and the observed dispersion of about $50\,\rm km/s$. For $C \simeq 1$,
the
agreement is better than can reasonably be expected from our crude estimates.
For plausible values of our parameters the turbulence excited by the large rate
of
supernovae can indeed excite all the turbulence observed on the one hand, and
required by the effective viscosity deduced from our modelling of the cloud
velocities on the other hand.  This is a consistency check, not a proof.  We
conclude that this mechanism may explain the observed cloud velocity
dispersion.

\TITLEA{4.\ The accretion rate in our Galactic Center}%
The characteristic velocities ($v_{\rm turb}$) and the length scale s
($l_{\rm turb}$) then provide the scales for
the exchange of momentum, here angular momentum, and so define a turbulent
viscosity by the product of thickness of distribution (twice the scale height)
and turbulent velocity of order

$$\nu_t\,=\,300 \; \rm pc \;km / s\;.\eqno{20}$$
In order to be an effective viscosity, this turbulence must tap the momentum
in the interstellar medium, and that is concentrated in the clouds.  This is
achieved by noting, that it is long known (Mathewson and Ford 1970,
Appenzeller 1971) that magnetic fields permeate the interstellar medium, from
cloud to intercloud medium and so effectively couple both phases. In the
following, we assume the magnetic field to play an important r\^ole only
in this respect. Our results (Linden et al. 1993, Biermann et al. 1993)
are in good agreement with the assumption that for the large scale dynamics in
the range between about ten and a few hundred parsec, the magnetic field does
not play the dominant r\^ole. Thus, in the following, we neglect its effect
on the large scale accretion flow.

The physics of such an accretion disk are better thought of in the
original accretion disk theory developed first by L\"ust (1952) than in the
picture of Shakura and Sunyaev's (1973) $\alpha$ disk models. A (not
necessarily constant) $\alpha$ value corresponds to a certain radial
variation of the effective viscosity that is directly coupled to the
thermal structure of the disk. This means that in $\alpha$ disk models, the
chosen value of $\alpha$ determines the thermal and mechanical structure of the
disk in the fashion of a closed system (with the only coupling to the
outside through radiative losses) while in our models we envisage
the value of the effective viscosity to be determined by processes other than
simple gas and/or dust friction in an inherently  turbulent shearing media (see
sect. 3). These processes themselves then are not necessarily directly
coupled to the accretion flow. In $\alpha$ disks, there is a limiting value
for $\alpha$ of the order of 1 (the exact value depends on the averaging
procedure for the vertical structure). Larger values correspond to
either supersonic (i.e., $v_{\rm turb} > c_s$) or to anisotropic
turbulence ($l_{\rm turb} > H$),
both of which are excluded in a standard
$\alpha$ disk. For our effective viscosity that is determined from
processes outside the closed gas/dust disk system, such a limit for
the viscosity does not apply as the driving process does not know about
$c_s$ and $H$ directly.

Given this viscosity the accretion velocity can be deduced from the
conservation of angular momentum and the assumption of hydrodynamic equilibrium
in vertical direction:

$$v_r \;=\;-  x \, {\nu_t \over s}  \eqno{21}$$
$x$ is a quantity of order 1; its actual value again depends on details of the
averaging procedure in vertical direction; for our estimates assuming
$x = 1$ suffices.
We reach the limit of validity of the accretion disk picture
when this implied radial velocity approaches
a noticeable fraction of the circular velocity, say $1/3 \,v_{\phi}$.  This
limit is reached at a radius of

$$r_{\rm crit}\;=\;({3 \over {v_{\phi,o}}} \,{\nu_t \over r_0})^{8/9} \,r_0.
\eqno{22}$$
where $v_{\phi,0}$ is the circular velocity at radius $r_0$.  Using the
parameters adopted we obtain for the critical radius $5.8$ pc, very close to
the
inner ring observed near $2$ pc. Again within the framework of our
approximations
this is a good agreement as we have to realize that we have deduced $r_{\rm
crit}$
neglecting any effects from the disk's inner boundary (Duschl and Tscharnuter,
1991). For $\Sigma \sim
s^{-1}$ and $M \sim r^{5/4}$, we find that effects due to the disk's
inner boundary scale $\sim s^{-7/8}$, i.e., become small only for radii $\ga
10 \rm pc$.

Since the critical radius very nearly scales
linearly with the turbulent viscosity, we might argue that the presence of this
ring implies that the turbulent viscosity is approximately a factor of $3$
smaller than used above near the radius of this ring.  Using then our first
estimate for the turbulent viscosity again of $300$ pc km/s the accretion
time scale

$$\tau_{\rm acc} \;=\; {r^2 \over \nu_t}  \eqno{23}$$
amounts to $2.8 \,10^5$ yrs at $r_0$ and $3.2 \,10^7$ yrs at $100$ pc.

A check can be made on the concept of overall accretion:  The accretion flow, a
tight spiral, influences cloud motion.  In fact, clouds can be used as tracers
of the flow field.  In Linden et al. (1993) we have followed one of the biggest
clouds, M-0.13-0.08 and its vicinity,
using it as a tracer, and were able to verify that this accretion disk
flow gives an excellent overall fit to the velocity field of this
large cloud and also implies independently, by the fit, a turbulent viscosity
of similar order of magnitude as derived here (about $2000$ pc km/s).  The
fit gives a distance of the cloud, at $115$ pc from the Galactic Center.

All these arguments suggest that the turbulent viscosity rises with radius in
an approximately linear fashion, from somewhere near $100$ pc km/s near $2$
pc
to about $2000$ pc km/s near $100$ pc.  It follows that the ratio of radial
velocity
$v_r$ to circular velocity $v_{\phi}$ is nearly independent of radial distance
$r$ in the disk, i.e. $s$.  The structure of the accretion flow is thus
approximately self-similar with the surface indeed going as $1/s$
asymptotically.  In this simplified model our earlier assumption of $\Sigma
\sim 1/s$ is thus justified.  Elsewhere we derived the radial dependence of
the effective viscosity from the observations of Zylka (1990) and Pauls et al.
(1993) and compared
it with a simple analytical model, where surface density and the
supernova excitation are checked self-consistently (Biermann et al. 1993).

We note in passing, that a turbulent viscosity of similar order of magnitude is
also implied at larger distances from the center in our Galaxy, where the cloud
motions are smaller, but the cloud distribution scale height is larger; in the
solar neighborhood the implied turbulent viscosity is of order $700$ pc km/s.
This implies that even on scales of order kpc, the accretion time scale is
still
of order $10^9 \,\rm yrs$, comparable to the timescale of galactic evolution
and reaching the Hubble time near at least several kpc.  On more conceptual
grounds, this has been argued already by Silk and Norman (1981), Lin and
Pringle (1987a, b), and Yoshii and Sommer-Larsen (1989).  Lin and Pringle
(1987b)
argue that self-gravity is the motor.  A detailed new calculation of galactic
evolution in a viscous disk allowing for the possibility of a central mass
concentration was performed by Diewald (1992). This in turn implies that
somewhere between $100$ pc and a few kpc the turbulent viscosity may reach a
maximum and then decrease again.

The observed star formation inside $500$ pc is in the range $0.3$ to $0.6
\, M_{\odot} \,\rm yr^{-1}$ (G\"usten 1989); with the assumption again,
that the surface density scales approximately with $1/s$, we thus obtain
for the observed star formation rate within $100$ pc $0.06$ to $0.12 \,
M_{\odot} \,\rm yr^{-1}$.  We note that the implied molecular gas
consumption time scale is to within a factor of $2$ uncertainty given by $3
\,10^8 \,\rm yr$, consistent with the estimate of $8 \,10^8 \rm yr$ for the
inner
disk of the galaxy M33 (Wilson et al. 1991).  Using the estimated surface
density of the gas in the central region given earlier the accretion rate
implied is of order $1/3 \; M_{\odot}\rm / yr$, which matches easily the
requirement that the observed star formation be fed.

\TITLEA{5.\ The energetics of driving both turbulence and star formation}%
Using the above analysis we can make another consistency check, and ask,
whether
the hot intercloud medium can actually be held at its temperature by the steady
energy input by supernovae. We note first that with our approximations [from
eqs.\
(16) and (18)]

$$n_{\rm intercloud} \;=\; 52 \,{\rm cm^{-3}\ } C^{-3} \,
(f_{\rm cl}/0.1)^{-1} \, (s/r_0)^{-1.21}, \eqno{24}$$
and [from (2) and (18)]

$$H \;=\; 1.1 \, 10^{19} \,{\rm cm\ } C \, (s/r_0)^{0.65} \,. \eqno{25}$$

The cooling in energy emitted per unit of time and area of the disk then is
given
by

$$\eqalign{&(1 - f_{\rm cl}) \,(10^{22} \, L) \, n_{\rm intercloud}^2 \, H
\;=2.7 \, 10^{37} { \rm erg/s/pc^2} \cr & \;\;\;C^{-5} \, (1 - f_{\rm cl})
\, (f_{\rm cl}/0.1)^{-2} \, (s/r_0)^{-1.77} \, .  }  \eqno{26}$$
This has to be balanced by the energy input from supernova explosions in our
model.  This we estimate by calculating the star formation rate from the
surface
density with a time scale of star formation proportional to the accretion (to
within a factor $\epsilon$), and using the entire galaxy with an estimated
supernova rate of one per $30$ years and a star formation rate of $10 \,\rm
M_{\odot} \, /year$ as reference point (Cox and Mezger, 1987). This leads then
from the adopted surface density law to an energy input of

$$1.3 \, 10^{39} \,{\rm erg/s/pc^2\ } \epsilon^{-1} \, C^2 \,
(s/r_0)^{-2.57} \,.\eqno{27}$$
This demonstrates first, that the energy dissipation in the intercloud medium
can
be balanced by the supernova activity.  Second, it also shows, that there is an
outer radius, where the energy balance fails.  This cutoff radius $s_{\star}
/r_0$
is at

$$\eqalign{&s_{\star}/r_0 \;=\; \cr & \, 29\cdot\left({{(f_{\rm cl}/0.1)^2}
\over {1 -
f_{\rm cl}}}\right)^{1.26} \, C^{8.82} \, \epsilon^{-1.26} \, (3 10^{21} \,
L)^{-1.26}.} \eqno{28}$$
For the parameters $f_{\rm dis}$ and $f_{\rm cl}$, the dependence is
$${s_\star \over r_0} \sim {{f_{\rm dis}^{5.70} f_{\rm cl}^{1.48}}\over
{(1 - f_{\rm cl} )^{1.26}}}.\eqno{29}$$
Here we have to check several things:  First, the parameter $f_{\rm dis}$
has to be small compared to unity, since otherwise there would not be any
shock;
we adopt $f_{\rm dis} = 0.1$.  Second, the parameter $f_{\rm cl}$, the cloud
volume filling
parameter, could be nearly anything $<1$; the data analysis of the cloud
motions
suggests that the high effective viscosity disk extends to at least $115$ pc,
and
so within our parameter range this is indeed consistent with $f_{\rm cl} \sim
0.1$.
Third, we have adjusted the cooling function to the higher value valid near
temperatures just below $10^5$ K.

Here, we can then ask again, whether the filling factor
of the medium can actually
reach unity, i.e. whether the filling factor of the intercloud
medium depends only weakly on radial distance $r$ and remains near unity over
the
radial range we consider.  With the expressions introduced above we have to add
an expression for the radial dependence of the star formation rate; we
assume that the radial dependence of the star formation rate is given by the
gas
surface density divided by the accretion time scale to within a factor of order
unity, since the accretion provides the fresh material for star formation.
Since the time scalesy may very well
be a function of the radius, this does not imply that we assume the star
formation rate to be linearly
proportional to the surface density $\Sigma$. We rather
replace the time derivative $\dot\Sigma_{\rm SF}$ by the difference quotient
$\Sigma /\tau_{\rm acc}$.

It is then easily seen that the filling factor indeed remains near unity over
the radial range considered; for a star formation rate which is more weakly
dependent on $r$, the radial dependence of the filling factor is even weaker.
Hence we have shown, that there is a self-consistent mode of highly excited
star formation.

The observations demonstrate that the Toomre
criterion is not marginally fulfilled, and the theoretical analysis above
suggests
that we have two limiting physical states of such a disk, one in which the
Toomre
instability starts clumping and thus can initiate star formation, and then a
higher level of activity when the turbulence is supernova driven, thus
providing
for a high effective viscosity, thus a high accretion rate, thus sufficient
material for further star formation and a sustained high rate of supernovae.
This
higher state of activity may exhaust the available supply of gas provided from
further out, and so may lead to an extreme limit cycle of star formation
activity
in the inner region of galactic disks with a consequent extreme limit cycle in
the
supply of gas to the innermost regions.  We have made several consistency
checks on this concept.
\vfill
\eject
\TITLEA{6.\ The feeding of AGN in analogy to our Galactic Center}%
If, as seems rather likely, the inner regions of galaxies, which in
observable contrast to our Galactic Center do harbor an active nucleus, show at
least all those complications that our Galactic Center demonstrates, then such
an
accretion rate is clearly sufficient to fuel also nuclear activity.  Since the
observation that the FIR/mm spectrum of radioweak active galactic nuclei (Chini
et
al.\ 1989a, 1989b; Sanders et al.\ 1989; Lawrence et al.\ 1991) is in all
likelihood dominated by dust emission on radial scales up to about $100\,\rm
pc$,
we also know that molecular clouds do exist -- this also has been confirmed
in a
number of examples (Sanders \& Scoville 1988, Barvainis \& Antonucci 1989,
Barvainis et al.\ 1989) -- and so that sufficient gas is available for
accretion.

All our above estimates and checks do not depend on properties that are
characteristic of our Galactic Center only. Thus the above deduced limit cycle
is most likely working in all galactic centers, especially in active ones.

\TITLEA{7.\ Conclusion and outlook}%
In many galaxies, bars are prominent features, and their action may very well
be the dominant mechanism
to remove angular momentum  from material that is many hundreds of pc
to a few kpc from the
galactic center. This will allow the material to be accreted into regions
closer to the center. But our investigations demonstrate that, on length scales
of the
order of tens to a few hundreds of pc, supernovae driven turbulence is capable
of inducing a sufficently large viscosity that allows for accretion of large
amounts of matter even in the absence of a bar.

It is interesting to speculate that the other transport coefficients important
for a galaxy and its evolution might be related to the transport coefficient
for angular momentum:  The transport coefficient that determines the leakage
of Cosmic Rays from a galaxy, as well as the transport coefficient that
enters the dynamo mechanism in order to produce the rather strong magnetic
fields which we observe.  Apart from arguments about the numerical value of
these transport coefficients one first consequence would be that we ought to
seriously consider the possibility that these other transport coefficients are
strong functions of radial distance in a galaxy.  It is also likely that any
theory for, say, a dynamo working in a galaxy cannot be separated from the
overall accretion flow (see Chiba and Lesch 1993), since both mechanisms depend
on
closely related microphysics.
\par
Our Galactic Center demonstrates that accretion via angular momentum transport
through turbulent viscosity is sufficiently large to provide all the gas
necessary in our Galactic Center to drive the star formation, and, in analogy,
to drive the accretion towards a central compact object in active galactic
nuclei.  The physical mechanism which we propose to drive the turbulence --
supernovae -- naturally leads to extreme limit cycles in the accretion rate.


\BEGREF{References}
\REF Appenzeller I., 1971, A\&A 12, 313
\REF Barvainis R., Alloin D., Antonucci R., 1989, ApJ 337, L69
\REF Barvainis R., Antonucci R., 1989, ApJS 70, 257
\REF Berkhuijsen E.M., 1986, A\&A 166, 257
\REF Bieging J.H., Abbott D.C., Churchwell E.B., 1989, ApJ 340, 518
\REF Biermann P.L., Duschl W.J., Linden S.v., 1993, A\&A (in press)
\REF Blitz, L., Binney, J., Lo, K.Y., Bally, J., Ho, P.T.P., 1993, Nature 361,
     417
\REF Bloemen H. (ed.), 1991, The interstellar disk-halo connection
     (IAU-Symp.\ 144)
\REF Bloemen J.B.G.M., Blitz L., Hermsen W., 1984, ApJ 279, 136
\REF Chevalier, R., 1974, ApJ 188, 501
\REF Chiba, M., Lesch, H., 1993, A\&A (submitted)
\REF Chini R., Kreysa E., Biermann P.L., 1989a, A\&A 219, 87
\REF Chini R., Biermann P.L., Kreysa E., Gem\"und H.-P., 1989b, A\&A 221, L3
\REF Cox D.P., 1972, ApJ 178, 159
\REF Cox P. and Mezger P.G., 1987, in: Star formation in galaxies (ed.: C.J.
     Lonsdale Perrson), NASA Conf.\ Publ.\ No. 2466, p.23
\REF Dickel, J.R., Sault, R., Arendt, R.G., Matsui, Y., Korista, K.T., 1988,
     ApJ 330, 254
\REF Diewald M., 1992, Masters Thesis, University of Bonn
\REF Duschl W.J., Tscharnuter W.M., 1991, A\&A 241, 153
\REF Falcke H., Biermann P.L., Duschl W.J., Mezger P.G., 1993,
     A\&A 270, 102
\REF Garcia-Munoz M., Mason G.M., Simpson J.A., 1977, ApJ 217, 859
\REF Genzel R., Townes C.H., 1987, ARA\&A 25, 377
\REF G\"usten R., 1989, in: The Center of the Galaxy, IAU Symposium No.\ 136,
     Ed.\ M.\ Morris, Kluwer, Dordrecht, p.\ 89
\REF Jackson J.M., Geis N., Genzel R., Harris A.I., Madden S.C., Poglitsch A.,
     Stacey G.J., Townes C.H., 1993, ApJ 402, 173
\REF Kronberg P.P., Biermann P., Schwab F.R., 1985, ApJ 291, 693
\REF Lawrence A., Rowan-Robinson M., Efstathiou A., Ward M.J., Elvis M., Smith
     M.G., Duncan W.D., Robson E.I., 1991. MNRAS, 248, 91
\REF Lebrun F. et al., 1983, ApJ 274, 231
\REF Lin D.N.C., Pringle J.E., 1987a, ApJ 320, L87
\REF Lin D.N.C., Pringle J.E., 1987b, MNRAS 225, 607
\REF Linden S.v., Duschl W.J., Biermann P.L., 1993, A\&A 269, 169
\REF Long, K.S., Blair, W.P., Van den Bergh, S., 1988, ApJ 333, 749
\REF L\"ust R, 1952, Zeitschr.\ f.\ Naturf.\ 7a, 87
\REF Markevitch M., Sunyaev R.A., Pavlinsky, M., 1993, Nature 364, 40
\REF Mathewson D.S., Ford V.L., 1970, MNRAS 74, 139
\REF McKee C.F. and Ostriker J.P., 1977, ApJ 218, 148
\REF Morris M., 1993, ApJ 408, 496
\REF Pauls T., Johnston K.J., Wilson T.L., Marr J.M., Rudolph A., 1993,
     ApJ 403, L13
\REF Reynolds R.J., 1989, ApJ 339, L29
\REF Reynolds R.J., 1990, ApJ 349, L17
\REF Sanders D.B., Scoville N.Z., Soifer B.T., 1988, ApJ 335, L1
\REF Sanders D.B., Phinney E.S., Neugebauer G., Soifer B.T., Matthews K., 1989,
     ApJ 347, 29
\REF Sanders R.H., Lowinger T., 1972 AJ 77, 292
\REF Schaaf R., Pietsch W., Biermann P.L., Kronberg P.P., Schmutzler T., 1989,
     ApJ 336, 722
\REF Schmutzler T., Tscharnuter W.M., 1993, A\&A 273,318
\REF Shlosman I., Begelman M.C., Frank J., 1990, Nature 345, 679
\REF Shlosman I., Begelman M.C., Frank J., 1990, Nature 345, 679
\REF Shlosman I., Frank J., Begelman M.C., 1989, Nature 338, 45
\REF Shakura N.I., Sunyaev R.A., 1973, A\&A 24, 337
\REF Silk J., Norman C., 1981, ApJ 247, 59
\REF Smith L.F., Biermann P., Mezger P.G., 1978, A\&A 66, 65
\REF Strom, R.G., Goss, W.M., Shaver, P.A., 1982, MNRAS 200, 473
\REF Sunyaev, R.A., Markevitch, M., Pavlinsky, M., 1993, ApJ 407, 606
\REF Toomre A., 1964, ApJ 139, 1217
\REF Wilson C.D., Scoville N., Rice W., 1991, AJ 101, 1293
\REF Yoshii Y., Sommer-Larsen J., 1989, MNRAS 236, 779
\REF Zylka R., 1990, Ph.D. Thesis, University of Bonn
\REF Zylka R., Mezger P.G., Wink J.E., 1990, A\&A 234, 133
\ENDREF

\bye